\documentstyle[11pt,aaspp4]{article}

\begin{document}

\title{Measuring the Diffuse Optical Light in Abell 1651}

\author{Anthony H. Gonzalez\altaffilmark{1}, Ann
I. Zabludoff\altaffilmark{2,3}, Dennis Zaritsky\altaffilmark{2,3}, and
Julianne J. Dalcanton\altaffilmark{4} }

\altaffiltext{1}{Department of Astronomy and Astrophysics, University
of California at Santa Cruz, Santa Cruz, CA 95064}
\altaffiltext{2}{University of California Observatories/Lick
Observatory and Department of Astronomy and Astrophysics, University
of California at Santa Cruz, Santa Cruz, CA 95064}
\altaffiltext{3}{Steward Observatory, University of Arizona, 933
North Cherry Avenue, Tuscon, AZ 85721} 
\altaffiltext{4}{Department of Astronomy, University of Washington,
Box 351580, Seattle, WA 98195-1580}

\begin{abstract}

 Using drift scan data, a new approach to determining surface
brightness profiles, and techniques for detecting low surface
brightness signals, we fit the light profile of the brightest cluster
galaxy (BCG) in the rich cluster Abell 1651 out to 670 $h^{-1}$
kpc. This radius is a significant fraction of the virial radius of the
cluster (2 $h^{-1}$ Mpc; \cite{gir98}), indicating that the sizes of
the BCG and the cluster are comparable. We find that the profile is
consistent with a de Vaucouleurs profile over the radial range probed.
We also find that the integrated light profile of the BCG in Abell
1651 contributes 36\% of the total cluster light within 500 $h^{-1}$
kpc. Including all luminous components, we obtain $M/L_I\sim$160 $h$
for the cluster, which would be overestimated by $\sim$20\% without
the BCG halo.  Furthermore, the relatively red color of the BCG at
large radii suggests that recent disruption and tidal stripping of
spirals and dwarf ellipticals do not contribute significantly to the
halo luminosity.  The color and the form of the profile are consistent
with a scenario in which the BCG forms from filamentary collapse
during the epoch of cluster formation, with relatively little
evolution in the past 5 Gyrs (\cite{dub98}). We remove the BCG and
other detected galaxies from the image and construct a two dimensional
surface brightness map of the cluster core. Several knots of excess
emission are found, but the total diffuse component is constrained to
contribute less than 5\% of the cluster light.

\end{abstract}


\clearpage

\section{Introduction}
Measuring the surface brightness profiles of brightest cluster
galaxies (BCGs) out to large radii is critical both for understanding
the formation of these giant galaxies and for determining the
mass-to-light ratios ($M/L$) of galaxy clusters. First, the form and
color of the profiles yield information about the dynamical state and
the distribution of stellar populations
(\cite{mal84,mer84,sch88,and95,dub98}), thus constraining the
accretion history of these systems.  Second, measurement of the light
contributed by the extended profile of the BCG is required for an
accurate determination of the total cluster luminosity, which is
essential for deriving an unbiased $M/L$.

Despite the importance of accurate BCG profile measurements, recent
determinations disagree. For example, from a large sample of BCGs,
Schombert (1986) finds that some of these galaxies have a `cD halo'
--- an extended component, centered on the BCG, with a significantly
different surface brightness profile than the central region of the
galaxy.  In contrast, Graham et al. (1996) find that BCGs can be fit
with single Sersic (1968) profiles, with profiles that are typically
shallower than a de Vaucouleurs ($r^{1/4}$) law. Meanwhile, the few
detailed analyses of diffuse light in individual clusters find BCGs
that are well-described by $r^{1/4}$ law profiles. Uson, Bough, \&
Kuhn (1990, 1991) show that the radial profile of the brightest
cluster galaxy in Abell 2029 is consistent with a de Vaucouleurs model
(reduced $\chi^2$=0.78 in $R$) out to $r\equiv(ab)^{1/2}=$425 $h^{-1}$
kpc, and Scheick \& Kuhn (1994) conclude that the BCG in Abell 2670 has an
r$^{1/4}$ profile (reduced $\chi^2$=0.28 in $V$) out to 230 $h^{-1}$
kpc.  The latter result is surprising given that this giant elliptical
is classified by both Oemler (1973) and Schombert (1986) as a cD
galaxy with a pronounced envelope starting at r$\sim$80 $h^{-1}$ kpc.

To unambiguously constrain BCG formation and permit accurate $M/L$
determination, we aim to resolve this disagreement in the form of BCG
luminosity profiles. A new study is needed that extends the detailed
analysis techniques used in Abell 2029 and 2670 to a statistical
sample of clusters.  However, attaining the required flatfielding
accuracy with pointed CCD observations is a computationally and
observationally intensive task (\cite{gud89,uso91,sch94}).  In this
paper we develop a method of studying the diffuse light that minimizes
the required telescope time and can be used to efficiently study a
large sample of clusters.  We employ very flat drift scan data, a new
approach to determining surface brightness profiles, and techniques
for detecting low surface brightness signals that, having been
developed for finding high-redshift clusters (\cite{dal96,zar97}), are
also applicable to this problem.

With these tools, we perform a detailed analysis of the distribution
of light in the cluster Abell 1651 at z=0.084.  We choose this cluster
to be our first because it appears to be dynamically relaxed, and
hence is a good system with which to test our method. Specifically,
X-ray observations show a symmetric temperature profile co-centric
with the brightest cluster galaxy, with a mean temperature T$_x=6.1$
keV (\cite{mark98}).  We focus on measuring the surface brightness
profile of the BCG, constraining the luminosity contribution of
diffuse light, and assessing the relative contributions of various
components to the total cluster luminosity.

\section{Data and Preliminary Reductions}

   Drift scan data were obtained using the Las Campanas 1m telescope,
the Great Circle Camera (\cite{zar96}), and the Tek\#5 CCD with both
the Gunn $i$ filter (transformed to Cousins $I$ using Landolt
standards) and a wide band filter (hereafter denoted as $W$) that
roughly covers the wavelength region between $B$ and $I$ (see Figure
1). This $W$ filter is designed to maximize the incident signal while
avoiding sky emission lines in the red and atmospheric refraction in
the blue.  Individual drift scans are 2048x13000 pixels with a plate
scale of 0.7$^{\prime\prime}$ pixel$^{-1}$ and an effective exposure
time of 95 seconds.  The total time required for a single scan,
including the time spent off-source, is $\sim$10 minutes. We have
three scans of the cluster core in $I$ and two in $W$, for total
exposure times of 4.75 minutes and 3.17 minutes.  Conditions during
the observations were photometric, with seeing of
1.5$^{\prime\prime}$.

 A key property of the data is the intrinsic flatness of drift
scans. The data must have residual flatness variation less than 0.5\%,
or these variations will be the dominant source of noise in our
derived surface brightness map.  In drift scans, pixel-to-pixel
variation is minimized as data are clocked across the chip, so
sensitivity variations are a concern only perpendicular to the readout
direction (at a level $\sim$ 2\% in our raw data).  Consequently, we
construct a one dimensional flatfield, for which the Poisson noise is
reduced significantly relative to a two-dimensional
flatfield. Flatfielding is accomplished in two stages. In both stages
we use a set of 6 ($W$) or 7 ($I$) data scans, to construct a median
averaged flatfield. Each scan is 13,000 pixels in length and the
typical sky level in $I$ is $\sim$140 counts, so the associated
Poisson noise is 0.03\% per column. The first stage of flatfielding
immediately follows bias subtraction and reduces sensitivity variation
from 2\% to 0.2\%. This level is below the noise from other sources;
however, the remaining variation due to the presence of objects in the
scans is correlated across columns. The second flatfielding stage is
designed to remove this residual variation. To eliminate contamination
from resolved objects, we use the segmentation image generated by
SExtractor version 2.0.15 (\cite{ber96}) to generate a binary mask.
We convolve this binary mask with a boxcar filter to mask all pixels
within 7$^{\prime\prime}$ of object detection regions. The second
stage flatfield image is constructed using only the unmasked
pixels. Subsequent to this final flatfielding, all scans are flat to
$<$0.1\%, which corresponds to $\mu_I$=28.4 mag arcsec$^{-2}$, and any
residual column-column variation is not discernible.

Similar to the sensitivity variation, temporal sky variability is a
one dimensional problem with drift scan data. This variability is a
smooth feature with maximum amplitude of $\sim$10\% of the sky level
over the length of a scan.  We apply a median boxcar of size
700$^{\prime\prime}$ (1000 pixels or 855 h$^{-1}$ kpc) to a
flatfielded version of each image in which all resolved objects are
masked. We find that a filter of this size does not cause
oversubtraction of the sky near the cluster core or bright stars. The
entire region within 350$^{\prime\prime}$ (500 pixels) of the cluster
core is also masked as a precaution.  We also try an alternative
method in which the sky is fit with a spline of order 5. Comparison of
the two methods yields rms variations at a level of $\mu_I\sim28$ mag
arcsec$^{-2}$, and uncertainty at the level of $\mu_I\sim30.5$ mag
arcsec$^{-2}$ in fitting the profile of the brightest cluster galaxy.

Following bias subtraction, flatfielding, and sky subtraction, we
register all images. The $I$- and $W$-band data are averaged to
generate a single image for each band and we also add the data from
both bands to maximize the signal available for tracing the BCG
profile at large radii. Adding the images also reduces uncertainty
from removal of the time variable sky component.

\section{The Cluster Components}

 To identify and characterize all significant sources of luminosity in
the core of the cluster, we proceed as follows. First, we model the
brightest cluster galaxy to determine the form of its surface
brightness profile (\S3.1). Next, we remove the brightest cluster
galaxy and all other detected objects from the image and analyze the
distribution of light from diffuse matter and faint, undetected
cluster galaxies (\S3.2). Finally, we use this information to assess
the relative contribution of each of these components to the total
cluster luminosity (\S3.3).

\subsection {The Brightest Cluster Galaxy}
We model the brightest cluster galaxy using the IRAF routine {\it
ellipse}.{\footnote{IRAF is distributed by the National Optical
Astronomy Observatories, which are operated by the Association of
Universities for Research in Astronomy, Inc., under cooperative
agreement with the National Science Foundation.}} SExtractor is again
used to generate an object mask.  Previous work has demonstrated that
contamination from other sources can significantly bias the derived
profile at low surface brightness levels (\cite{uso91,por91}), and so
we mask all objects except the BCG, and all pixels within
10$^{\prime\prime}$ of object detection regions.  This procedure
eliminates 30\% of the image data from further consideration by {\it
ellipse}. The masking is then augmented by manual masking of regions
near saturated stars and bright galaxies, which excludes an additional
15\% of the pixels in the image.{\footnote{The regions of excess
surface brightness described in \S 3.2 are also masked at this point,
and so are not responsible for any observed structure in the
profile.}} Our approach is similar to that utilized by
Vilchez-G\'{o}mez, Pell\'{o}, \& Sanahuja (1994) in their study of the
diffuse light of Abell 2390.  We generate best fit models for the BCG,
with position angle and ellipticity first allowed to vary freely, and
then with these parameters fixed to the values given by SExtractor.
The resultant surface brightness profile, as a function of
$r=(ab)^{1/2}$, is unchanged whether these parameters are fixed or
allowed to vary.  We choose to fix both parameters.

To measure the surface brightness profile of the BCG at large radii,
an accurate determination of the sky level is critical. Figure 2
illustrates the effect on an intrinsic de Vaucouleurs profile of error
in determination of the sky level.  Such an error in the background
leads either to the truncation of the profile or to the existence of
an artificial envelope.  As has been noted by a number of authors
(\cite{dev70,oem76,mel77,uso91}), systematic errors in the measured
sky level can be induced by intrinsic background variation, spatial
variation in detector response (e.g. flatness variation for CCDs), and
contamination from the outer halos of other cluster members.  For this
data set we calculate that the uncertainty in our sky determination is
0.01\% ($\sigma_{sky}\sim29.5$ mag arcsec$^{-2}$ in $I$), which will
impact the form of the profile at $r>200 h^{-1}$ kpc.  To
unambiguously quantify the form of the profile at larger radii, we
must employ a technique that is insensitive to level of the background
light.

Our alternate method uses the differential change in the flux, $\Delta
f$, between points in the profile.  By taking a differential
measurement, we obtain a quantity that is independent of a constant
background level, but still contains all the information present in
the luminosity profile.  One difficulty exists with this approach. If
we compute the differential change only for radially adjacent surface
brightness measurements ($\Delta f_i \equiv f_i-f_{i+1}$), then both
the reduced $\chi^2$ ($\chi^2_v$) and the error bars for the model
parameters are dependent on radial sampling density.  Higher sampling
density leads to lower $\chi^2_\nu$ and larger error bars on the model
parameters because the flux difference between adjacent points
decreases, while the associated uncertainty does not.

Our preferred method, described in the Appendix, is to compute for
each point the flux difference relative to all points at larger
radii. We define
\begin{equation} 
\Delta f_i\equiv f_i - \frac{1}{N-i}\sum_{j=i+1}^{N} f_j,
\end{equation}
and compare this quantity with model predictions.  This approach
removes the dependence of $\chi^2_\nu$ upon sampling density,
permitting robust determination of both model parameters and their
associated uncertainties.

Using {\it{ellipse}}, we measure the profile of the BCG in Abell 1651
out to $r$=670 $h^{-1}$ kpc. Beyond this radius, the angular extent of
the elliptical annulus used to measure the profile exceeds the width
of the image.  We show in Figure 3{\it{a}} the flux differential in
$I$+$W$ for the BCG in Abell 1651, and our best fit de Vaucouleurs
model.  A best-fit Sersic (1968) model and a model with a cD envelope
are also shown in the residual plot, Figure 3{\it{b}}.{\footnote{The
properties of the cD envelope are equivalent to the envelope observed
by Schombert (1988) for Abell 2670. It is modelled with a de
Vaucouleurs profile with $r_e$=330 $h^{-1}$ kpc and
$\Sigma_e=0.008\Sigma_{e,galaxy}$. }} The de Vaucouleurs model fit to
the differential profile has $r_e=(41.7\pm0.8) h^{-1}$ kpc (68\%
confidence) with $\chi_\nu^2$=13.7.  The large $\chi_\nu^2$ is due to
radial oscillations in the light profile (see Figure 3{\it{b}}), which
is seen in both filters.  This radial structure is not due to using a
fixed position angle and eccentricity; the same oscillations are seen
when these parameters are permitted to vary.  The best-fit Sersic
model formally has $n=4.3\pm0.2$ with $\chi^2_\nu=14.3$, which is
slightly shallower than a de Vaucouleurs model. However, the
oscillations seen in the residuals may contribute to this result.  For
example, if the three points beyond 300 $h^{-1}$ kpc (which appear to
be on a rising part of the oscillation pattern) are excluded from the
fit, then the best-fit model has $n$=3.9, and so we conclude that the
Sersic index of the galaxy profile is consistent with n=4 (de
Vaucouleurs) to within our observational uncertainty.  We also note
that the cD envelope model shown in Figure 3{\it{b}} yields
$\chi_\nu^2$=21.6.

Figure 4 shows the $I$+$W$ composite surface brightness profile of the
BCG in Abell 1651, with the background level fixed using the results
from the differential analysis, and the vertical scale set such that
$\mu$(1 $h^{-1}$ kpc)=0 mag arcsec$^{-2}$ (see also Table 1). The de
Vaucouleurs, Sersic and cD envelope models shown are the same as in
Figure 3.  A fit to the stellar PSF is also overlaid for comparison,
demonstrating that seeing has a negligible impact on the
profile. Figure 5 shows the $I$- and $W$-band profiles independently,
and de Vaucouleurs fits, with the effective radius fixed to $r_e$=41.7
$h^{-1}$ kpc.  Error bars represent observed rms flux variations in
the data and do not include systematic errors.  For $r_e$=41.7
$h^{-1}$ kpc, independent fits in the $I$- and $W$-bands respectively
yield $\mu_e(I)$=23.55 mag arcsec$^{-2}$ with $\chi_\nu^2$=9.4, and
$\mu_e(W)$=24.70 mag arcsec$^{-2}$ with $\chi_\nu^2$=14.7.

In addition to measuring the form of the profile, we also test for the
presence of a color gradient.  While we are able to measure the
profiles out to $r\sim$500 $h^{-1}$ kpc in both bands, for $r>$100
$h^{-1}$ kpc we are wary of systematic errors that may bias the
resultant colors. Consequently, we restrict our attention to the inner
100 $h^{-1}$ kpc.  We find a mild gradient in the profile
($\Delta(W-I)/(\Delta\log r)=0.25\pm$0.08) from 15-100 $h^{-1}$ kpc,
with the halo at 100 $h^{-1}$ kpc redder than the center of the BCG by
0.2 mag. This mild gradient is consistent with other studies that have
found shallow or no color gradients in BCGs (\cite{mac92,gar97}). For
comparison, in their study of 17 non-BCG elliptical galaxies
(-22.5$<$M$_B<$-20), Franx, Illingworth, \& Heckman (1989) find in the
mean a slightly blue radial gradient ($\Delta(B$-$R)/(\Delta\log
r)\sim-0.1$).

The color gradient of the BCG is a potentially valuable probe of the
evolutionary history of the system. Qualitatively, if significant
recent accretion of cluster galaxies has occurred in the halo, then we
should expect the halo to be blue relative to the core of the BCG,
independent of whether the accreted systems are spirals or
ellipticals.  Spirals are bluer because of ongoing star formation,
while the relative blueness of fainter ellipticals is primarily a
metallicity effect (\cite{lar74,kau98,fer99}). For reference, we
estimate that if one $\sim L_*$ spiral ($\sim$5\% of the total
luminosity of the BCG) is accreted within the past 2 Gyrs and
deposited uniformly at 50$<$r$<$100 $h^{-1}$ kpc, the halo at these
radii will be $\sim$0.1 mag bluer after accretion.  Conversely, for
formation scenarios in which there is no significant, recent
contribution to the outer halo from tidally disrupted systems, we
expect either no gradient, or a mild red gradient if there has been
subsequent star formation in the center of the galaxy.  One such
formation scenario is demonstrated by Dubinski (1998) in a simulation
that also produces a de Vaucouleurs profile for the
BCG. Unfortunately, quantitative model predictions are currently
lacking, and so, while our observations qualitatively agree with the
early formation scenario, further modelling is needed for
confirmation.

\subsection{The Cluster Surface Brightness Distribution}

Is there a discernible presence of intracluster light that is not
associated/co-centric with the BCG? To investigate this issue, we
first subtract from the image the BCG using our model from \S 3.1, and
all other galaxies with $I$-band isophotal areas larger than 175
pixels using the IRAF package GIM2D (\cite{sim98,mar98}).  GIM2D
generates optimal bulge+disk fits for each object, with an exponential
disk and de Vaucouleurs profile bulge. The motivation for this
detailed approach is to remove not only the visible central regions of
these galaxies, but also their contribution to the total cluster light
at fainter surface brightness levels.

Subsequent to the application of GIM2D, we use FOCAS
(\cite{jar81,val93}) to remove all remaining detected objects with
m$_I<$21 (or m$_W<22.8$) and replace them with randomly drawn, local
sky pixels. The interior regions of the previously removed bright
galaxies are also replaced by locally drawn random sky pixels as a
precaution against residual galactic light contaminating the cluster
surface brightness map. At this stage, we also mask bright stars and
the inner 35$^{\prime\prime}$ of the BCG.  Next, we convolve the
cleaned image with a 10$^{\prime\prime}$ (12.2 $h^{-1}$ kpc)
exponential kernel to generate a smoothed two dimensional map of the
core region of the cluster. This kernel size is chosen as a compromise
between sensitivity and resolution, being sufficiently small to
resolve individual cluster galaxies, but wide enough to probe to
$\mu>$26 mag arcsec$^{-1}$. Residual variations persist after
smoothing at an approximate rms level $\mu_I$=26.65 mag arcsec$^{-2}$,
and we detect fluctuations down to $\mu_I\sim$26 mag arcsec$^{-2}$.
Figures 6 illustrates the process used to generate the surface
brightness map. For comparison, the bottom panel, in which the
brightest cluster galaxy halo is not removed, is included.

We find that within a radius of 400 $h^{-1}$ Mpc of the BCG there
exist three regions of excess brightness with peak surface
brightnesses\footnote{Note that this is the peak surface brightness
after convolution. The true peak surface brightnesses could be
significantly brighter if the sources have spatial scales much smaller
than the smoothing kernel (i.e. individual faint galaxies).}
$\mu_I<25.75$ mag arcsec$^{-2}$. These regions are denoted as A, B,
and C in the middle right panel of Figure 6. An additional ring of
excess brightness can also be seen at the bottom of the panel
surrounding a masked star, but this ring is due to the extended point
spread function of the star. Other regions of excess brightness can
also be seen, such as the one between the BCG and the bright star, but
these fall below our $\mu_I=25.75$ mag arcsec$^{-2}$ threshold.  Of
the three peaks, B is the brightest and largest, with a peak surface
brightness $\mu_I$=25.3 mag arcsec$^{-2}$. The total excess luminosity
from this region corresponds to $L_I$=$(1.4\pm0.5)\times10^{10}
L_\odot$. Region B is aligned with the semimajor axis of the BCG (see
Figure 6).

To discern whether these enhancements are diffuse in origin, we look
for evidence of faint galaxies coincident with the bright regions. In
region B, the detected faint galaxy population cannot account for the
excess flux. Coupled with its large angular size, this eliminates the
possibility that the excess is due to a higher redshift group or
cluster.  This region is possibly the remnant of a tidally disrupted
cluster galaxy. Similar low surface brightness features have been seen
in Coma (\cite{gre98}), and can be explained as arising from tidal
processes.  Alternately, this region could arise from a significant
local enhancement of galaxies fainter than can be detected in our
data. This interpretation seems less plausible, because it would
require an excess relative to the expected average cluster density of
at least 20 galaxies fainter than our detection threshold, with no
corresponding overdensity of galaxies brighter than the detection
threshold.

In contrast to region B, inspection of regions A and C reveals a group
of faint galaxies associated with each surface brightness excess.
These galaxies have m$_I>$21 mag arcsec$^{-2}$, and so were not
removed from the image by FOCAS prior to smoothing. In both cases the
flux associated with the fluctuation can be accounted for by light
from these faint galaxies. Such groups could either be clumped dwarf
galaxies associated with the cluster, or more distant clusters or
groups.  Based on the work of Gonzalez et al. (2000) detecting distant
clusters, the surface density of background clusters is such that
$\sim$1 would be detected in this field, and so is consistent with the
latter explanation. To constrain the origin of fluctuations A and C,
we measure the $W-I$ color of the associated faint galaxies. These
galaxies are shown in Figure 7 and compared to the color-magnitude
relation for the cluster, derived using galaxies with m$_W<$20 (dashed
line). The weighted mean color of the ensemble is $W-I$=0.97
($\sim$0.3 mag bluer than the core of the BCG), which corresponds to
$R-I$$\sim$0.58. This color is consistent with the galaxies being
dwarf ellipticals in the cluster, and inconsistent with their being
giant ellipticals at higher redshift.  Based on color alone, however,
we cannot rule out the possibility that these galaxies are clustered
spirals at higher redshift. If these are indeed cluster dwarfs, it is
unclear why they are tightly clustered apart from any bright cluster
galaxies.

\subsection{Relative Luminosity Contributions}

In $\S 3.2$ we found that a de Vaucouleurs profile with a total
luminosity of $L_I=1.17\times10^{12} h^{-2}$ L$_\odot$ is a good fit
to the BCG light profile.{\footnote{The luminosity is corrected for
a galactic extinction of 0.05 magnitudes (\cite{schl98}).}}  To assess
the fractional contribution of the BCG to the total cluster light, we
examine the region within $r$=500 $h^{-1}$ kpc of the center of the
BCG. As mentioned earlier, the location of the BCG is
consistent with the kinematic center of the cluster as defined by
X-ray data.  Roughly 98\% of the luminosity of the
central galaxy is contained within this radius.  We calculate the
total luminosity from cluster galaxies with m$_I<$20.5 within the same
region. Statistical background subtraction is used to correct for the
contribution of galaxies not associated with the cluster. We use the
region of the image further than 2 $h^{-1}$ Mpc from the cluster core,
covering to an area of 0.58 sq. degrees, to compute the average
off-cluster contribution.  The integrated flux for galaxies with
$m_I<$20.5 is computed in both cluster and off-cluster regions, with
background subtraction resulting in a 25\% reduction in the observed
cluster flux.  The summed, extinction-corrected luminosity of cluster
galaxies, excluding the BCG, is $L_I=2.1 \times10^{12}$ L$_\odot$.

To estimate the luminosity of cluster galaxies or intracluster light
below this magnitude level, we use the surface brightness map (middle
right in Figure 6).  We cannot discern whether light in the surface
brightness map is due to faint galaxies or diffuse intracluster light,
but instead place an upper bound on the combined contribution by
summing the total residual light within the same $r$=500 $h^{-1}$ kpc
elliptical region and background subtracting using an off-cluster
region of the scan. To compute the mean off-cluster sky level, we use
a region adjacent to this ellipse that extends 3000 pixels
(35$^{\prime}$) in right ascension in both directions. The largest
source of uncertainty in this measurement is due to large scale
residual gradients in the background sky level, generated primarily by
scattered light from off-image stars. Including a conservative
estimate for this uncertainty, we calculate an excess flux of
110$^{+190}_{-110}$ counts sec$^{-1}$ within $r$=500 $h^{-1}$ kpc,
which corresponds to L$=(0.7^{+1.2}_{-0.7})\times10^{11} h^{-2}$
L$_\odot$, or 2$^{+3}_{-2}$\% of the total cluster light in this
region.  Because the error bars are large, we are not able to place a
meaningful constraint on the faint end of the luminosity function. All
faint end Schechter function slopes with $\alpha>-1.98$ for $m>$20.5
are consistent with the observed flux to within 3$\sigma$.  Our data
are shallow, but our observed luminosity function has an upturn at the
faint end in both bands and a slope $|\alpha|\sim 1.7 - 1.9$.  This
measurement suggests that any excess light can be explained as arising
from faint cluster galaxies.

Table 2 lists the total and fractional contributions of each component
of the net cluster luminosity within 500 $h^{-1}$ kpc.  The central
dominant elliptical is a key contributor to the cluster luminosity.
The BCG fractional contribution of 36\% is in good agreement with the
results from Uson, Bough, \& Kuhn (1990, 1991) for Abell 2029 and
Scheick \& Kuhn (1994) for Abell 2670, who find that the BCG's
contribute 23\% of the cluster light within 780 $h^{-1}$ kpc and
30$\pm$8\% of the total cluster light, respectively.  Furthermore, as
in Abell 2029, diffuse light beyond that associated with the BCG is a
negligible contributor to the total light of this system.

\section{The Mass-to-Light Ratio}

X-ray and optical spectroscopic data exist for Abell 1651, enabling us
to determine the cluster mass and hence the $I$-band mass-to-light
ratio.  Markevitch et al. (1998) computed an emission weighted
temperature of 6.1$\pm$0.4 keV (90\% confidence) for the
cluster. Girardi et al. (1998) compute a cluster velocity dispersion
$\sigma$=1006$^{+118}_{-96}$ km s$^{-1}$. Combining these two value
yields
\begin{equation}
\beta\equiv\frac{\mu m_p \sigma^2}{kT}=1.04\pm0.24
\end{equation}
for $\mu$=0.6, indicating that the gas and galaxies trace the
potential with the same energy per unit mass, and suggesting that the
cluster core is not far from equilibrium.{\footnote{Girardi et
al. (1998) claim that this cluster has significant substructure, in
conflict with the derived value of $\beta$ and detailed X-ray analysis
(\cite{mark98}).  However, this finding is based on the distribution
of a relatively small sample of 30 cluster galaxies.}}  To compute the
mass enclosed within 500 $h^{-1}$ kpc, we follow the method of Wu
(1994), which assumes an isothermal gas distribution. This
approximation is reasonable, as the temperature profile of Abell 1651
is quite flat (\cite{mark98}).  Evrard, Metzler, \& Navarro (1996)
demonstrated in their simulations that isothermal $\beta$ models give
an unbiased estimate of the true mass, but also demonstrated that
there is $\sim$20\% scatter in the relation between the computed and
true mass. With the isothermal assumption, the mass within a given
projected radius, $r_p$, is computed as a function of $\beta$, $T$, a
core radius $r_c$, and a `physical size' $R$ for the cluster, which
defines the line-of-sight depth over which to integrate the enclosed
mass. For a projected radius of 500 $h^{-1}$ kpc the derived mass is
only a weak function of $r_c$ and $R$. For example, changing $R$ from
3 $h^{-1}$ Mpc to 8 $h^{-1}$ Mpc modifies the resulting mass by 3\%,
while changing $r_c$ from 20 $h^{-1}$ kpc to 100 $h^{-1}$ kpc alters
the resulting mass by 2\%.  This uncertainty is significantly smaller
than the scatter from use of the isothermal model. For concreteness,
we adopt $r_c$=50 $h^{-1}$ kpc and $R$=5 $h^{-1}$ Mpc. Including the
scatter seen in the simulations, we compute
\begin{equation}
M(r_p<500 h^{-1} kpc) = (5.4\pm1.4) \times10^{14} h^{-1} M_\odot.
\end{equation}
The total $I$-band mass-to-light ratio within this region then is 
\begin{equation}
M/L_I(r_p<500 h^{-1} kpc) = (160\pm45) h.
\end{equation} 

A key point to note is the dependence of M/L$_I$ on the inclusion of
the extended BCG halo. Using only the BCG magnitude returned by
SExtractor, we would have missed 50\% of the light from the BCG, and
hence 18\% of the cluster light.  Consequently, if such extended halos
are a generic feature of brightest cluster galaxies, as they appear to
be from our work and that of Uson, Bough, \& Kuhn (1990, 1991) and Scheick
\& Kuhn (1994), then any M/L ratio that fails to account for this
light will also overestimate M/L by $\sim$20\%. This omission of
luminosity exacerbates the cluster baryon problem for high $\Omega_0$
models (\cite{whi93}).

\section{Formation of the BCG Halo}

 What do the properties of the BCG tell us about the formation history
of this system?  We have found that the surface brightness profile of
the BCG in Abell 1651 is, to first order, consistent with a de
Vaucouleurs model, and have observed that the profile becomes mildly
redder with increasing radius. The uniformity of the profile over such
a large radial range argues for early assembly of the extended
halo.{\footnote{The presence of oscillations in the profile also
provides a potentially interesting constraint, however, dynamical
modelling is first needed to assess the timescale over which such
structure can be maintainted in the cluster environment.}}  Meanwhile,
the color of the halo indicates recently accreted cluster galaxies do
not contribute a significant fraction of the halo luminosity, and so
also supports early formation.

These results are consistent with the recent work of Dubinski
(1998). In his hydrodynamic cluster simulation, the BCG is assembled
at z$\sim$0.8 and significant accretion continues until z$\sim$0.4,
after which there are no major mergers involving the BCG.  Dubinski
finds that a brightest cluster galaxy with an r$^{1/4}$ profile out to
200 $h^{-1}$ kpc is formed in this simulation via the merger of
massive galaxies during filamentary collapse.  He also notes that, as
we observe in Abell 1651, `a cD galaxy envelope did not form in this
system'.  In addition, the brightest cluster galaxy in the simulation
retains a fossil alignment with the filament and the galaxy
distribution.  Such an alignment with the galaxy distribution and
X-ray gas has been observed in real systems
(\cite{sas68,car80,por91,all95,mul98}). In Abell 1651 the BCG is
aligned with both the X-ray gas contours (Figure 8{\it{a}}) and, more
marginally, with the distribution of confirmed cluster members
obtained from NED (Figure 8{\it{b}}).  The alignment of Clump B with
the BCG can also be interpreted in this picture as the recent
disruption of a galaxy infalling along the direction of the filament.

\section{Summary and Conclusions}

We perform a detailed analysis of the distribution of luminous matter
in the galaxy cluster Abell 1651. We assess the relative luminosity
contributions of the brightest cluster galaxy, the rest of the cluster
galaxy population, and any diffuse, luminous matter in the cluster
that is unassociated with the other two components. In the process, we
develop and demonstrate an approach for studying the distribution of
luminous matter in cluster cores that allows detailed analysis, but
requires minimal telescope time ($<$ 1 hour per cluster on a
40$^{\prime\prime}$ telescope).  This technique can be applied to a
large sample of clusters to test whether there exists true variation
in the form of BCG surface brightness profiles, and also whether
intracluster light is a significant contributor in some systems.

In the case of Abell 1651, we find that the brightest cluster galaxy
contains 36\% of the total cluster light within $r$=500 $h^{-1}$ kpc
of the center of the BCG and the cluster.  Furthermore, the profile of
the BCG is well approximated by a de Vaucouleurs model out to $r$=670
$h^{-1}$ kpc, which is consistent with it being a D type galaxy
(\cite{mat64,mor75,sch87}).  Our current measurements extend further
in radius than those of Uson, Bough, \& Kuhn (1990, 1991) for the
cluster Abell 2029 and Scheick \& Kuhn (1994) for Abell 2670, and we
find close agreement with the results of these studies.  We also
detect a color gradient in the profile in the sense that the halo is
redder than the core of the BCG.  The properties of the brightest
cluster galaxy are consistent with the scenario of filamentary
collapse and formation of the BCG at high redshift, as evidenced by
Dubinski's (1996) numerical simulations.

We constrain the luminosity of any additional diffuse matter in the
 cluster to constitute less than $\sim$5\% of the total luminosity.
 Furthermore, the residual flux observed is consistent with arising
 from cluster galaxies fainter than the magnitude limit of our
 data. Thus, diffuse light beyond that of the BCG profile is
 negligible for this system. We do detect one distinct patch of excess
 light along the semimajor axis of the BCG (region B) that appears to
 be truly diffuse light, possibly from a tidally disrupted galaxy. Two
 additional, fainter peaks in the surface brightness are also seen,
 but are coincident with faint, clumped galaxies and have fluxes that
 are consistent with arising from these faint galaxies. The colors of
 these faint galaxies are consistent with the expected color of
 cluster dwarf ellipticals, and so a reasonable explanation is that
 these are groups of clumped cluster dwarfs. It is also possible that
 these galaxies are spirals in background clusters or groups.

Finally, we compute the $I$-band mass-to-light ratio and find
$M/L_I$=(160$\pm$45) $h$.  We note that failure to include the
luminosity of the BCG halo causes $M/L$ to be overestimated by
18\%. If this halo profile is typical of BCGs, any study of cluster
mass-to-light ratios that ignores this significant contribution to the
cluster light will systematically overestimate $M/L$. Omission of this
light will also induce a systematic bias in measurements of $\Omega_0$
derived using the resultant mass-to-light ratios.

\section{Acknowledgements}
We thank Carnegie Observatories for access to their facilities, and
Roelof de Jong, Eric Bell, and Hans-Walter Rix for helpful
discussions.  We also thank the referee, Jim Schombert, for helpful
comments and suggestions. AHG acknowledges support from the National
Science Foundation Graduate Research Fellowship Program and the ARCS
Foundation.  AIZ acknowledges support from NASA grant HF-01087.01-96A.
DZ acknowledges financial support from National Science Foundation
CAREER grant AST-9733111, and fellowships from the David and Lucile
Packard Foundation and Alfred P. Sloan Foundation.  JD acknowledges
support from NASA grant HF-01057.01-94A and GO-07327.01-96A. This
research has made use of the NASA/IPAC Extragalactic Database (NED)
which is operated by the Jet Propulsion Laboratory, California
Institute of Technology, under contract with the National Aeronautics
and Space Administration.

\appendix
\section{Appendix}
The standard technique for fitting surface brightness profiles is via
$\chi^2$ minimization of the surface brightness data using the
equation,
\begin{equation}
\chi_\nu^2 =\frac{1}{N-M-1} \sum_{i=1}^{N} \frac{(y_i - f_i)^2}{\sigma_i^2},
\end{equation}
where $\chi_\nu^2$ is the reduced $\chi^2$, $N$ is the number of data
points, $M$ is the number of free parameters, $y_i$ is the measured
flux at point $i$, $f_i$ is the value of the functional fit at point
$i$, and $\sigma_i$ is the uncertainty in the flux. For a de
Vaucouleurs model there are three free parameters, $r_e$, $\Sigma_e$,
and the sky level. Sersic models have $n$ as a fourth free parameter.
This approach is adequate for determination of $r_e$ and $\Sigma_e$;
however, fluctuations in the interior region of the galaxy
(particularly correlated fluctuations arising from small-scale
structure in the profile) have an inordinate impact on the resultant
background level. For example, in our data for Abell 1651,
determination of the sky level in this fashion leads to truncation of
the profile beyond $r\sim$ 300 $h^{-1}$ kpc, despite the fact that the
surface brightness is observed to monotonically decrease to beyond
$r$=600 $h^{-1}$ kpc. For Sersic models, error in the background sky
level also induces error in the derived $n$, and so this method
provides little leverage on the form of the profile at large radii.

To avoid the ambiguity that arises from uncertainty in the background
level, we have opted in this paper for a differential approach to
determining the form of surface brightness profiles. A quick way to
gain intuition for this approach is to plot the flux difference
between adjacent data points ($\Delta f_i = f_i - f_{i+1}$) as a
function of radius and compare this with various models.  However, as
discussed in the text, use of adjacent points has the drawback that
$\chi^2_\nu$, defined as
\begin{eqnarray}
\chi_\nu^2 =\frac{1}{(N-1)-M-1} \sum_{i=1}^{N-1} \frac{((y_i-y_{i+1}) 
- (f_i-f_{i+1}))^2}{\sqrt{\sigma_i^2+\sigma_{i+1}^2}},
\end{eqnarray}
is dependent upon the radial sampling density, and so it is not
possible to determine the uncertainty in the derived parameters.

Instead, for each point in the profile we compute the average flux
decrement between that point and all other points in the profile at
larger radii. Mathematically,
\begin{equation}
\Delta f_i\equiv f_i - \frac{1}{N-i}\sum_{j=i+1}^{N} f_j.
\end{equation}
This internal referencing allows us to compute a mean $\chi^2$ value
for each point in the profile, with larger separations receiving
larger weighting in determination of this mean value (because $\sigma$
is roughly constant, while the flux decrement increases). The
$\chi_\nu^2$ equation for this method is:
\begin{equation}
\chi_\nu^2 =\frac{1}{(N-1)-M-1} \sum_{i=1}^{N-1} \left[
\frac{1}{N-i}\sum_{j=i+1}^{N} \frac{((y_i-y_j) -
(f_i-f_j))^2}{\sqrt{\sigma_i^2+\sigma_j^2}}\right],
\end{equation}
with $\chi_\nu^2$ now independent of sampling density.  Because of the
increased leverage at large radii, this method permits a more robust
determination of $n$ than is possible with the standard surface
brightness fitting technique, while yielding comparable
values of $r_e$ and $\Sigma_e$.

\clearpage
\begin{deluxetable}{rrlll}
\tablecaption{ Surface Brightness Profile Data \label{tbl-1}}
\tablewidth{0pt}
\tablehead{
\colhead{$r$ \tablenotemark{a}} &\colhead{$r$} &\colhead{$I$+$W$ \tablenotemark{b}} & \colhead{$I$} &\colhead{$W$} \nl
\colhead{($^{\prime\prime}$)} &\colhead{($h^{-1}$ kpc)} &\colhead{(mag arsec$^{-2}$)} & \colhead{(mag arsec$^{-2}$)} &\colhead{(mag arsec$^{-2}$)} 
}

\startdata
  0.65 &   0.80 &  $-$0.09$\pm$0.04              &  18.58$\pm$0.04          & 19.72$\pm$0.04               \nl
  0.92 &   1.12 &  \hspace{0.3cm}0.05$\pm$0.04              &  18.71$\pm$0.05          & 19.85$\pm$0.04               \nl
  1.28 &   1.57 &  \hspace{0.3cm}0.27$^{+0.05}_{-0.04}$     &  18.93$\pm$0.05          & 20.07$\pm$0.04               \nl
  1.79 &   2.19 &  \hspace{0.3cm}0.58$\pm$0.03              &  19.24$\pm$0.03          & 20.38$^{+0.02}_{-0.03}$      \nl
  2.51 &   3.07 &  \hspace{0.3cm}0.95$\pm$0.02              &  19.61$^{+0.03}_{-0.02}$ & 20.76$\pm$0.02               \nl
  3.52 &   4.30 &  \hspace{0.3cm}1.32$\pm$0.02              &  19.98$\pm$0.02          & 21.12$\pm$0.01               \nl
  4.92 &   6.01 &  \hspace{0.3cm}1.69$\pm$0.01              &  20.34$\pm$0.01          & 21.5$\pm$0.01                \nl
  6.89 &   8.42 &  \hspace{0.3cm}2.08$\pm$0.02              &  20.74$\pm$0.02          & 21.88$\pm$0.02               \nl
  9.65 &  11.79 &  \hspace{0.3cm}2.58$\pm$0.02              &  21.24$\pm$0.02          & 22.38$\pm$0.02               \nl
 13.51 &  16.50 &  \hspace{0.3cm}3.15$\pm$0.02              &  21.80$\pm$0.02          & 22.96$\pm$0.02               \nl
 18.91 &  23.10 &  \hspace{0.3cm}3.81$\pm$0.03              &  22.47$\pm$0.04          & 23.61$\pm$0.03               \nl
 26.47 &  32.34 &  \hspace{0.3cm}4.57$\pm$0.02              &  23.18$\pm$0.03          & 24.42$\pm$0.02               \nl
 37.06 &  45.28 &  \hspace{0.3cm}5.13$^{+0.04}_{-0.03}$     &  23.74$\pm$0.04          & 24.98$\pm$0.03               \nl
 51.88 &  63.39 &  \hspace{0.3cm}6.10$^{+0.04}_{-0.03}$     &  24.69$\pm$0.06          & 26.01$\pm$0.04               \nl
 72.64 &  88.74 &  \hspace{0.3cm}6.67$\pm$0.05              &  25.24$^{+0.07}_{-0.06}$ & 26.59$\pm$0.06               \nl
101.69 & 124.24 &  \hspace{0.3cm}7.49$\pm$0.08              &  26.14$^{+0.11}_{-0.10}$ & 27.38$\pm$0.08               \nl
142.37 & 173.93 &  \hspace{0.3cm}8.25$^{+0.13}_{-0.12}$     &  26.79$^{+0.18}_{-0.15}$ & 28.21$^{+0.15}_{-0.13}$      \nl
199.31 & 243.51 &  \hspace{0.3cm}9.34$^{+0.33}_{-0.25}$     &  28.02$^{+0.50}_{-0.34}$ & 29.78$^{+0.73}_{-0.43}$      \nl
279.04 & 340.92 & \hspace{0.1cm}10.17$^{+0.83}_{-0.46}$     &  28.71$^{+1.47}_{-0.60}$ & 30.81$^{+\infty}_{-0.78}$ \nl
390.66 & 477.28 & \hspace{0.1cm}11.82$^{+\infty}_{-1.37}$& \hspace{-0.3cm}$<$29.55 \tablenotemark{c}  & 31.14$^{+\infty}_{-1.00}$ \nl
546.93 & 668.20 & \hspace{0.1cm}12.74$^{+\infty}_{-1.86}$&                          &				\nl

\enddata
\tablenotetext{a}{The seeing during these observations was 1.5$^{\prime\prime}$,
 and so data at smaller radii do not reflect the true profile of the galaxy.}
\tablenotetext{b}{The values given for $I$+$W$ have been arbitrarily normalized such that
$\mu$(1 $h^{-1}$ kpc)=0 mag arcsec$^{-2}$.}
\tablenotetext{c}{At this radius the $I$-band flux was indistinguishable from the background level, and so we are only
able to place a lower bound on the magnitude.}
\end{deluxetable}

\clearpage
\begin{deluxetable}{lrr}
\tablecaption{ Abell 1651 $I$-band Luminosity Budget \label{tbl-2}}
\tablewidth{0pt}
\tablehead{
\colhead{Source} &\colhead{$h^{-2} L_\odot$} & \colhead{Fraction of Light} 
}

\startdata
Brightest Cluster Galaxy 	 & $1.2\times10^{12}$  &  36\%     \nl	
Luminosity function(m$<$20.5)   	 & $2.1\times10^{12}$  &  62\%     \nl	
LF(m$>$21) + other diffuse light & $7\times10^{10}$ &  $2\%$   \nl   
\enddata
\end{deluxetable}

\clearpage
\centerline {\bf Figure Captions}
\bigskip

\figcaption[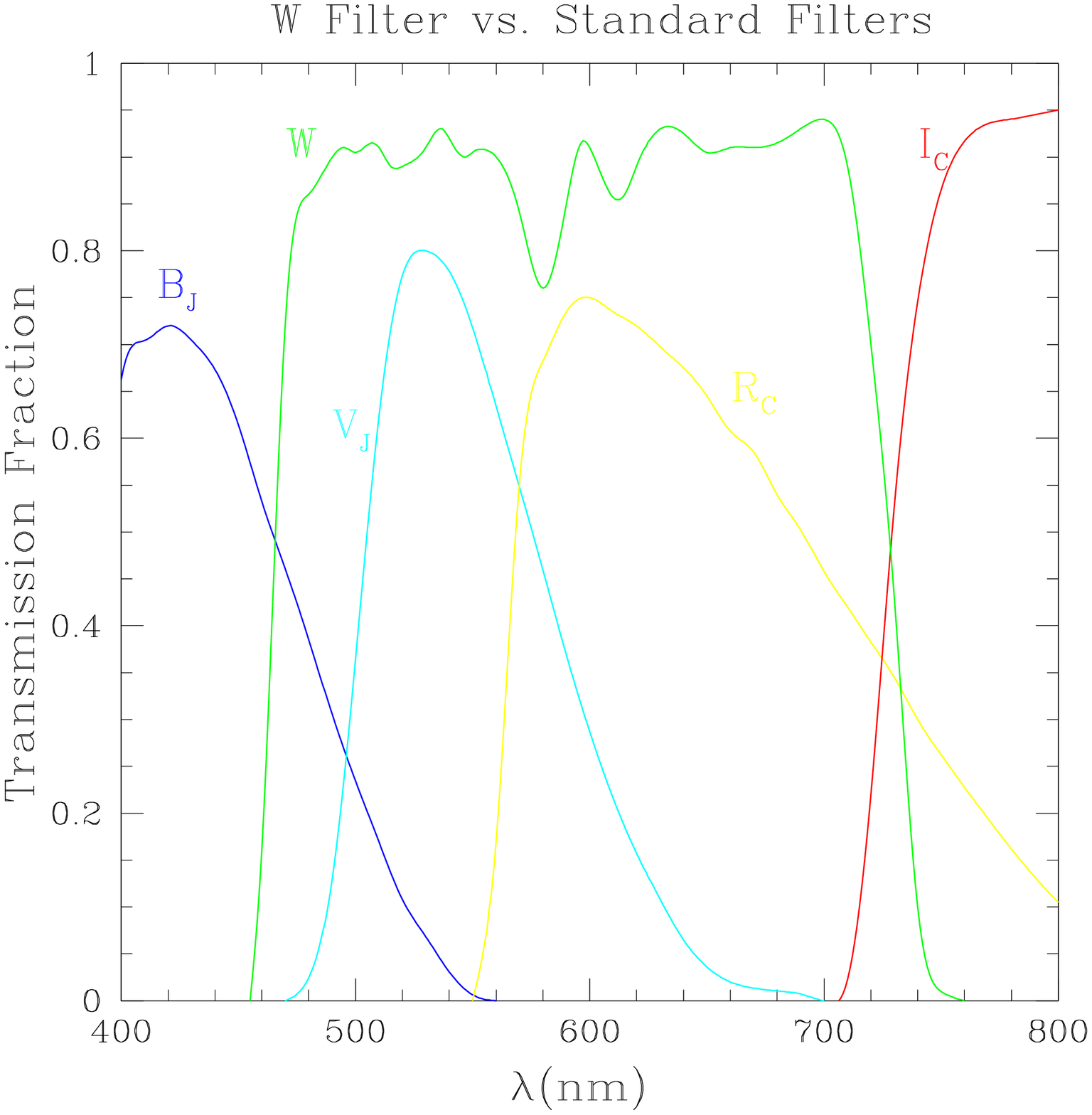]{Comparison of the broad $W$ filter utilized in
this work to standard Johnson ($BV$) and Cousins ($RI$) filters.  The
red cutoff is designed to avoid night sky lines, while maximizing
incident flux. \label{fig1}}

\figcaption[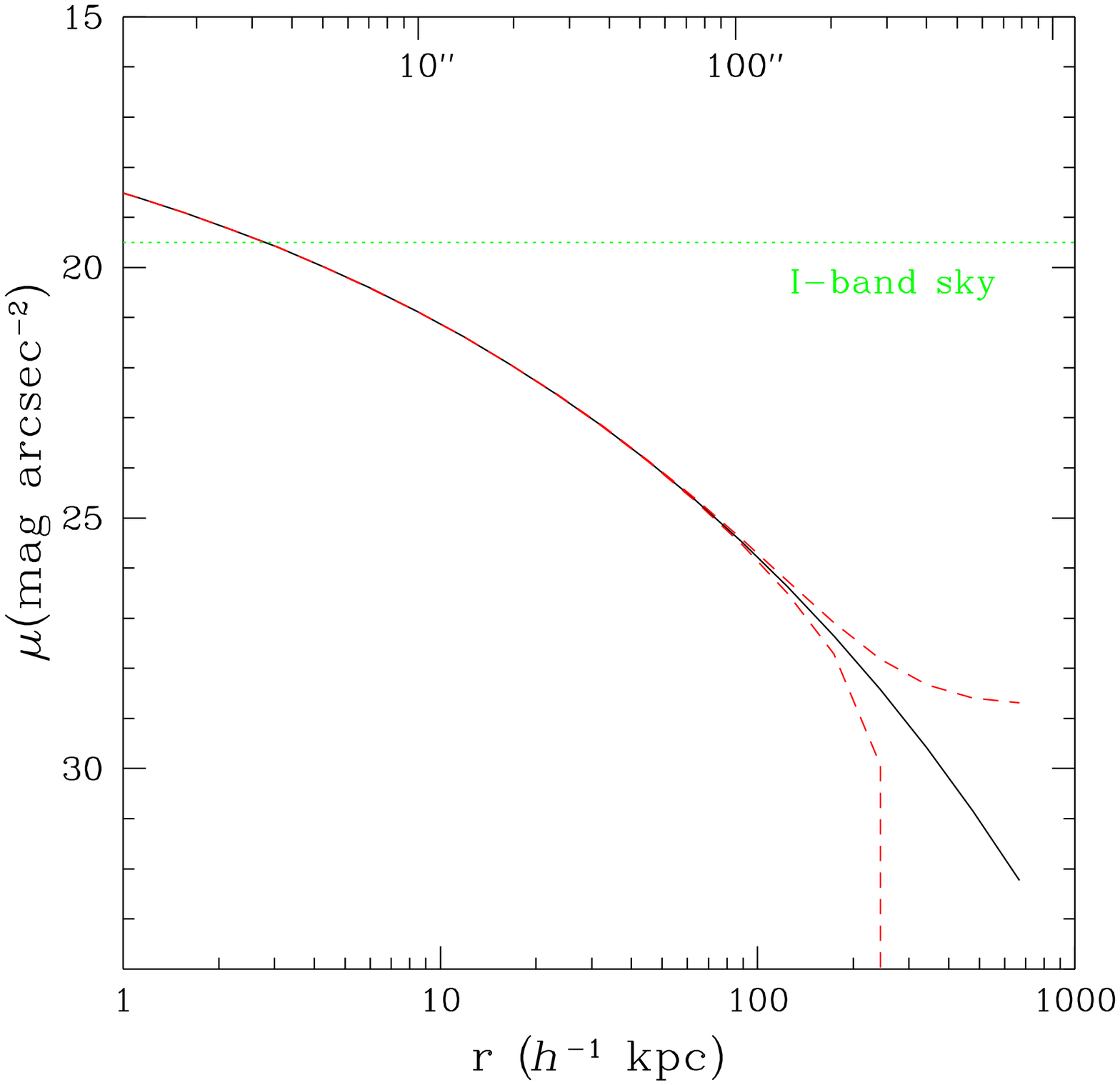]{The effect of a 0.01\% error in the sky level
(comparable to our observations) on an initial de Vaucouleurs profile.
Error in sky estimation can lead to either truncation of the profile
or an artificial excess of light. \label{fig2}}

\figcaption[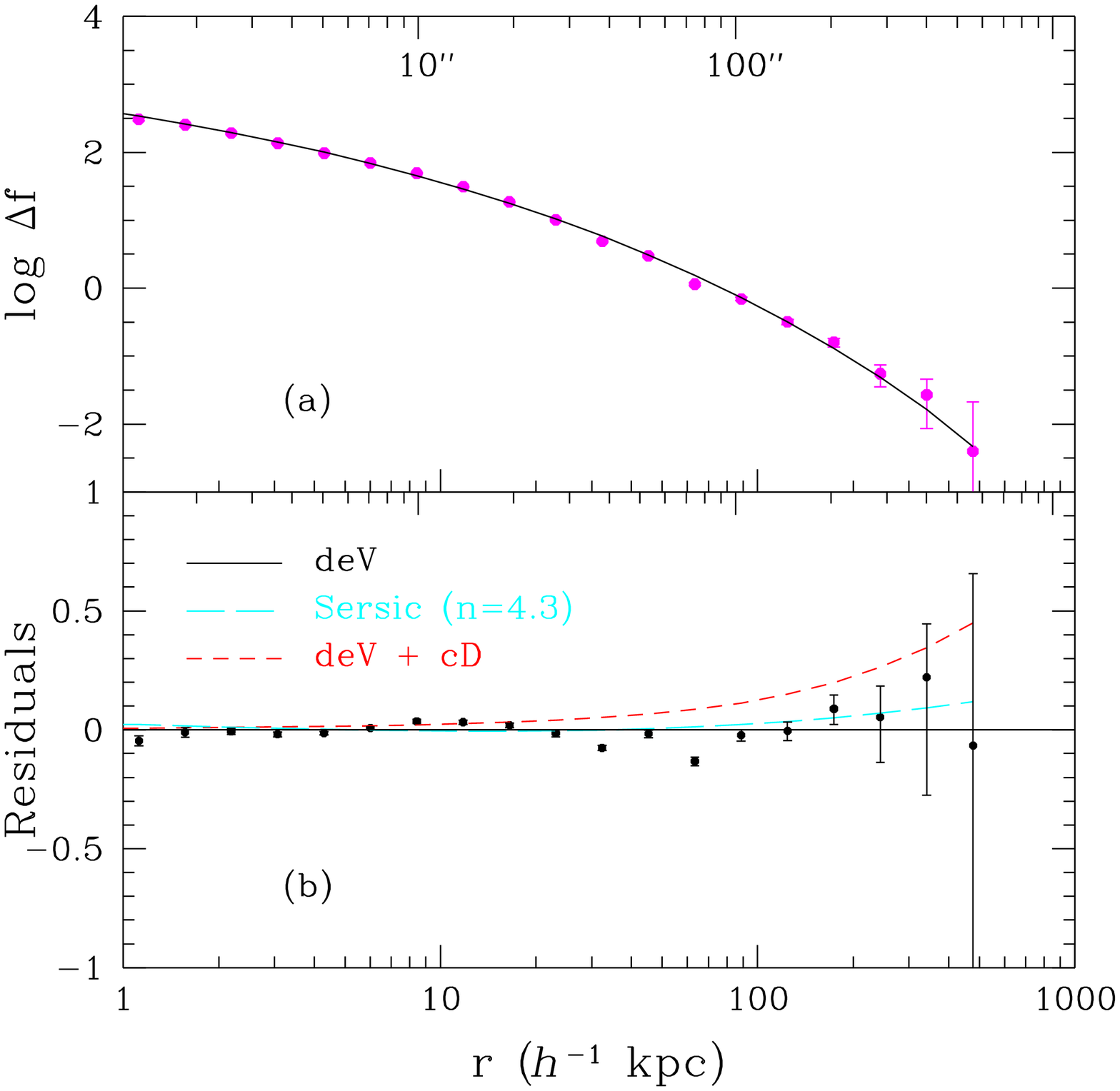]{(a) $\log \Delta f $ as a function of radius,
with $\Delta f$ defined as in equation (1).  The solid line is the
best fitting de Vaucouleurs model ($r_e$=41.7 $h^{-1}$ kpc).  (b)
Residual plot for the upper graph showing variation about the model
fit. Also shown are the optimal $n$=4.3 Sersic model (long dashed) and
a model with a de Vaucouleurs galaxy plus a de Vaucouleurs cD envelope
with $r_e$=330 $h^{-1}$ kpc and $\Sigma_{e}$=0.008$\Sigma_{e,galaxy}$
(short dashed).
\label{fig3}}

\figcaption[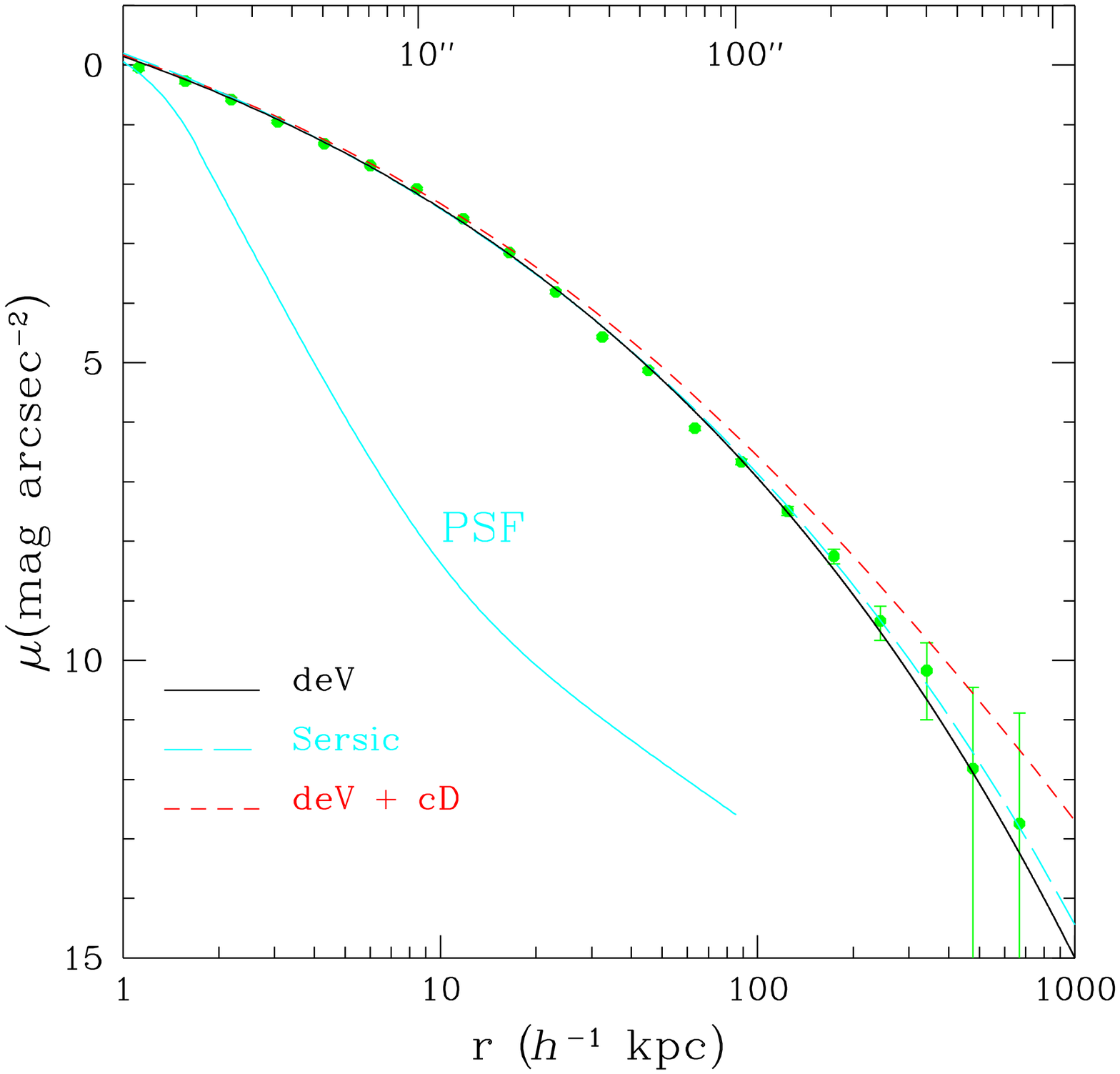]{Composite $I$+$W$ surface brightness
profile. The lines are the same models as in Figure 3. \label{fig4}}

\figcaption[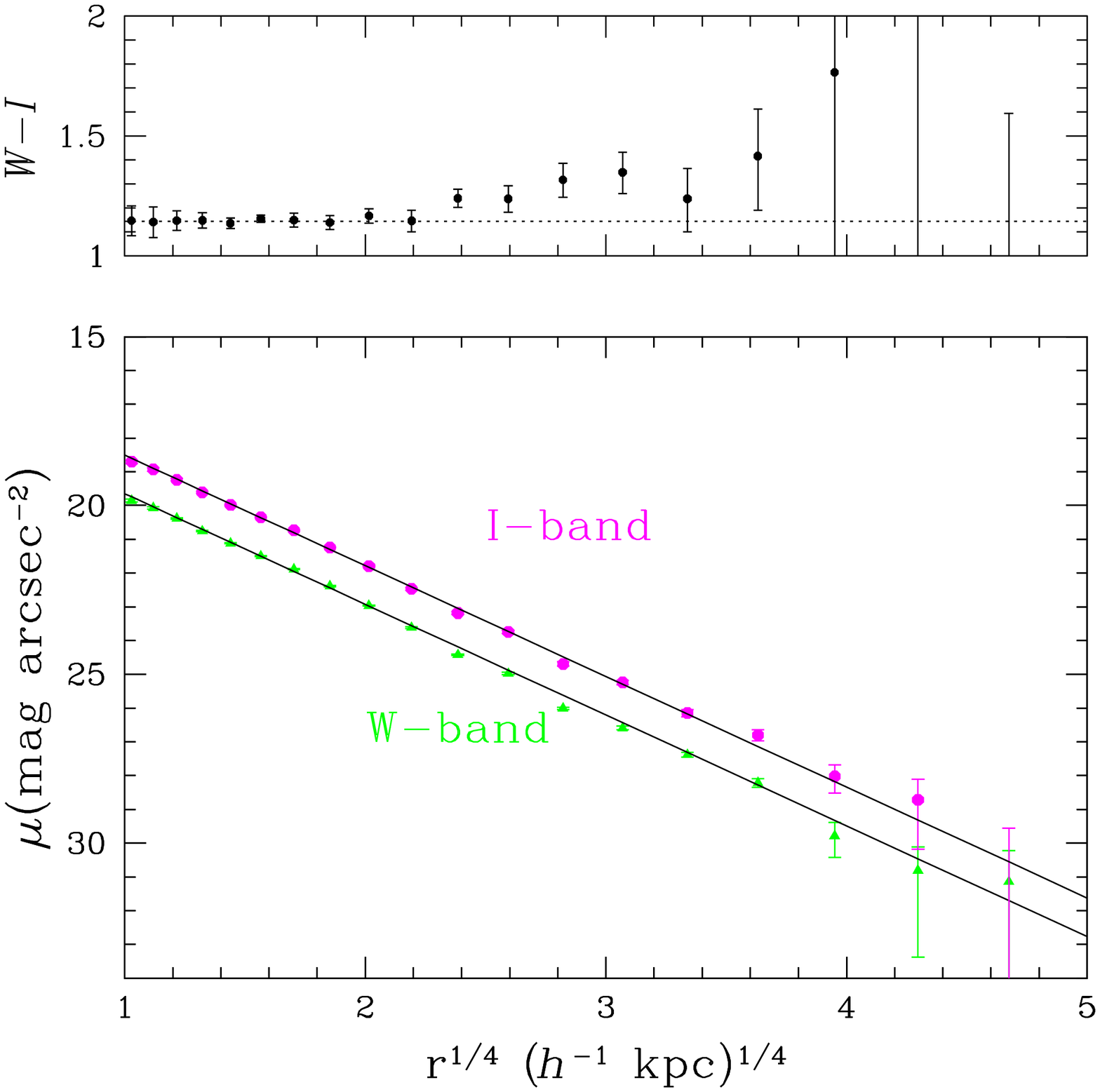]{Bottom: $I$-band (circles) and $W$-band
(triangles) surface brightness profiles. Solid lines are best fit
models as in Figure 4. Top: BCG color as a function of radius. A mild
reddening is evident from 20-100 $h^{-1}$ kpc.  At larger radii, the
derived color is sensitive to the background level, and so should be
viewed with caution.
\label{fig5}}

\figcaption[f6a.eps]{Sequence of $I-$band images illustrating the
process used to generate the surface brightness map. Starting with the
original image (upper left), first the BCG and brightest galaxies are
modelled and removed (upper right). Next, all objects with $\mu_I<21$
mag arcsec$^{-2}$ are replaced with locally drawn sky pixels and masks
are applied to the bright objects (middle left). Finally, this image
is convolved with an exponential kernel to generate the SB map (middle
right). The most significant feature in the surface brightness map is
the region marked B, which has a peak surface brightness of
$\mu_I=25.3$ mag arcsec$^{-2}$. For comparison, a surface brightness
map is also shown in which the halo of the bcg is not modelled and
removed (bottom). In all images, east is up and north is to the left.
Each frame is 700$^{\prime\prime}$ (855 $h^{-1}$ kpc) on a side.
\label{fig6}}

\figcaption[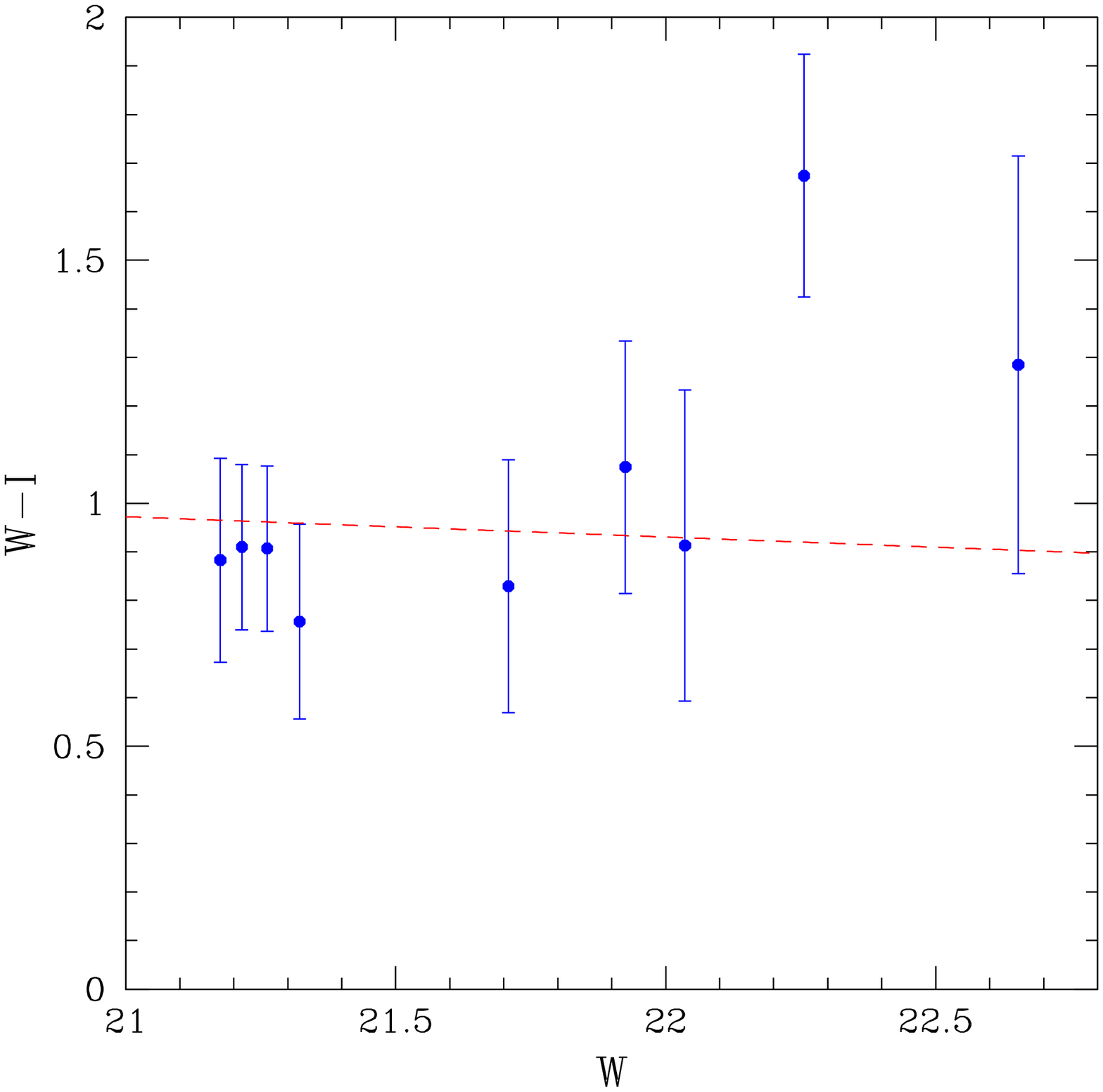]{Color-magnitude diagram for the faint galaxies
coincident with excess surface brightness regions A and C. The dashed
line is the red sequence for the cluster constructed from galaxies
with m$_W<20$. \label{fig7}}

\figcaption[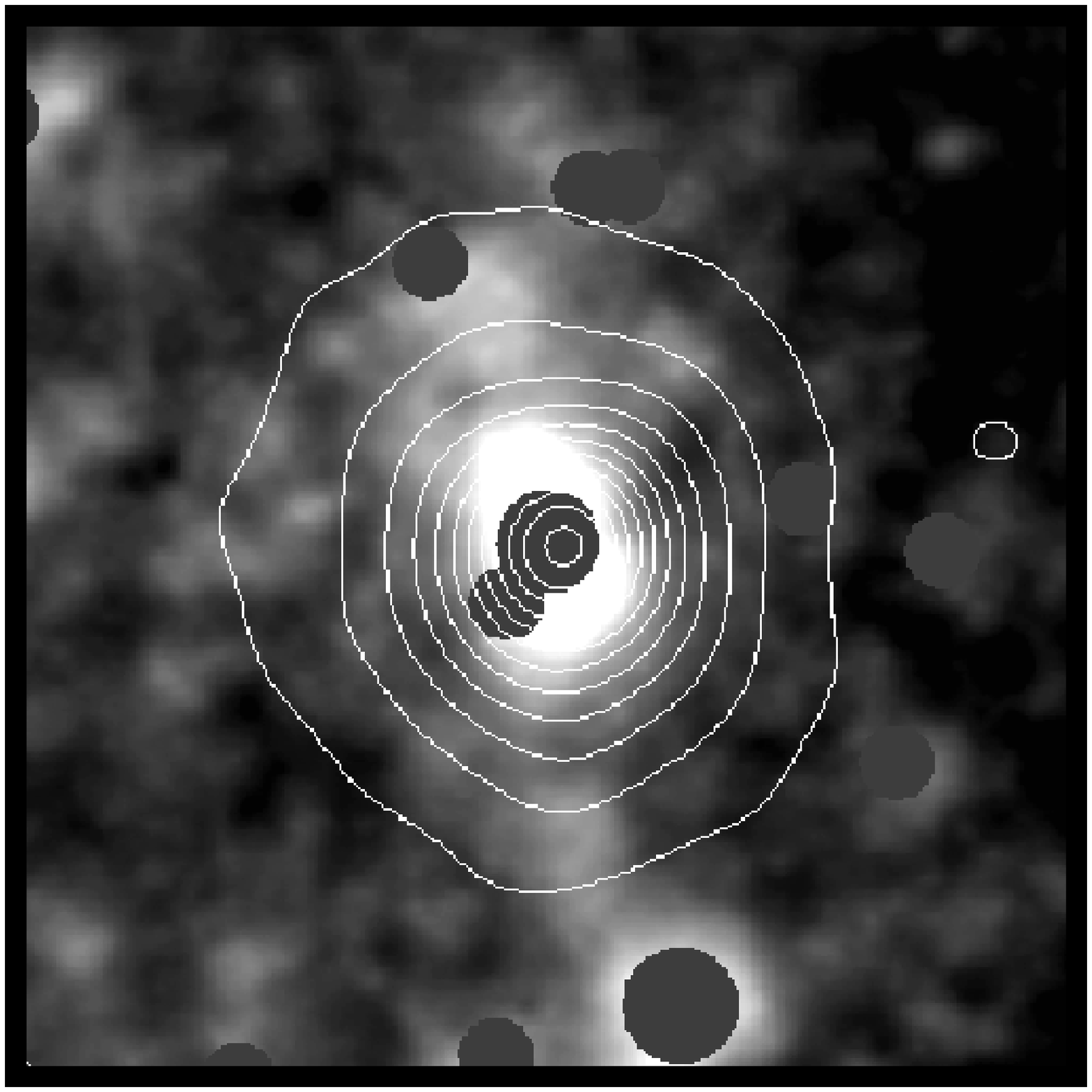]{ (a) X-ray contours overlaid on the smoothed
image of the BCG halo. As in Figure 6, the image is 855 $h^{-1}$ kpc
on a side. The X-ray image is from the ROSAT PSPC, with an exposure
time of 7435 sec.  To generate to contours shown, the image is
smoothed with a Gaussian of 30 arcsec FWHM; the resulting map has a
resolution of about 40$^{\prime\prime}$ (49 $h^{-1}$ kpc). This
resolution is determined fro the FWHM of several point sources lying
just outside the displayed field.  The lowest contour level is
3$-\sigma$ above the background and contour levels are displayed in
1$-\sigma$ increments.
(b)The distribution of all spectroscopically confirmed members of the
cluster that lie within a 15$^{\prime}$ (1100 $h^{-1}$ kpc) radius of
the BCG. The confirmed members preferentially represent the bright end
of the cluster luminosity function. The solid line indicates the
position angle of the BCG. The inner box denotes the region displayed
in Figure 8{\it{a}}.  East is up, north is to the left.
\label{fig8}}

\clearpage
\begin{figure}
\epsscale{1} \plotone{f1.eps}
\end{figure}
\clearpage
\begin{figure}
\plotone{f2.eps}
\end{figure}
\clearpage
\begin{figure}
\plotone{f3.eps}
\end{figure}
\clearpage
\begin{figure}
\plotone{f4.eps}
\end{figure}
\clearpage
\begin{figure}
\plotone{f5.eps}
\end{figure}
\clearpage
\begin{figure}
\plotone{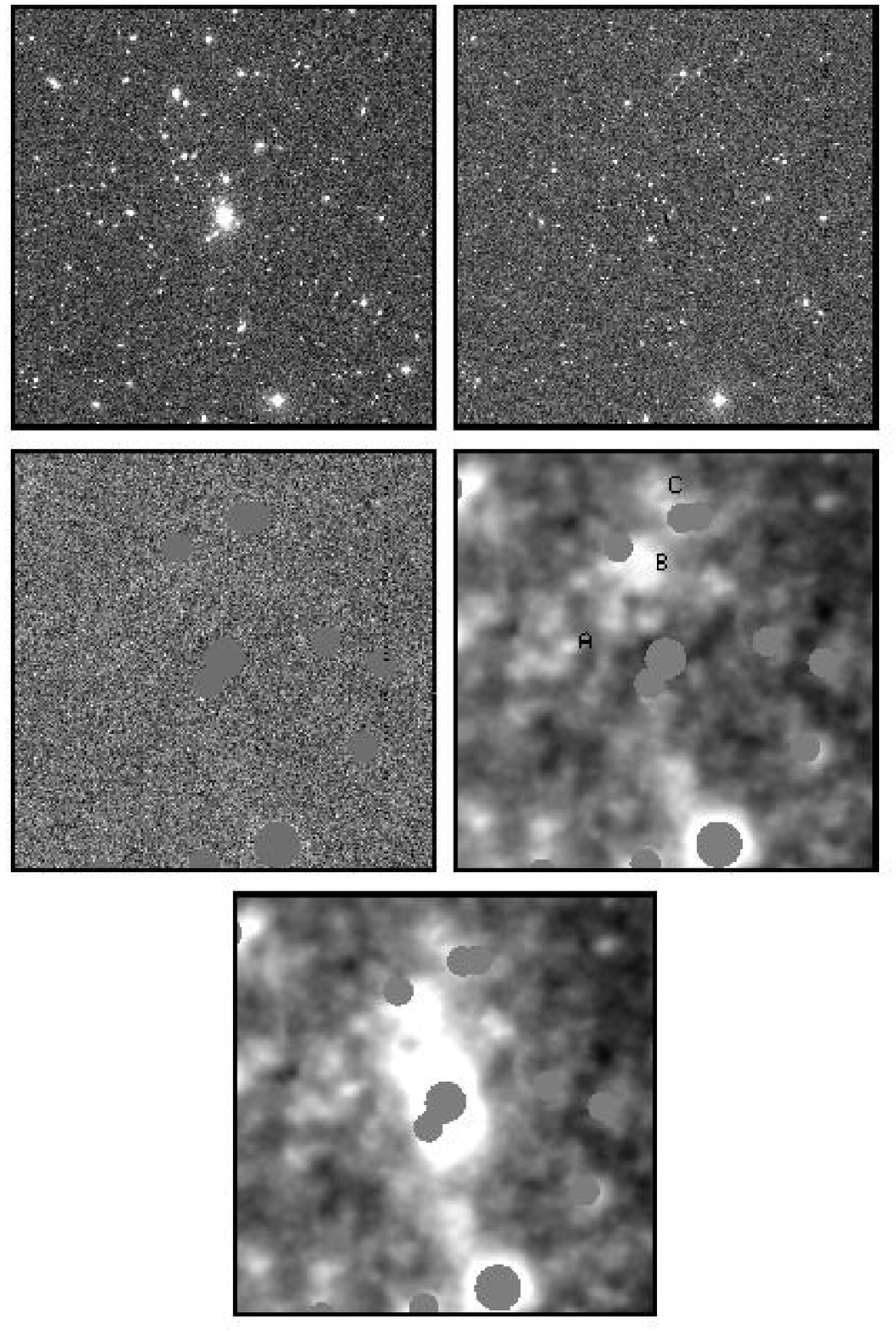}


\end{figure}
\clearpage
\begin{figure}
\epsscale{1}
\plotone{f7.eps}
\end{figure}
\begin{figure}
\plottwo{f8a.eps}{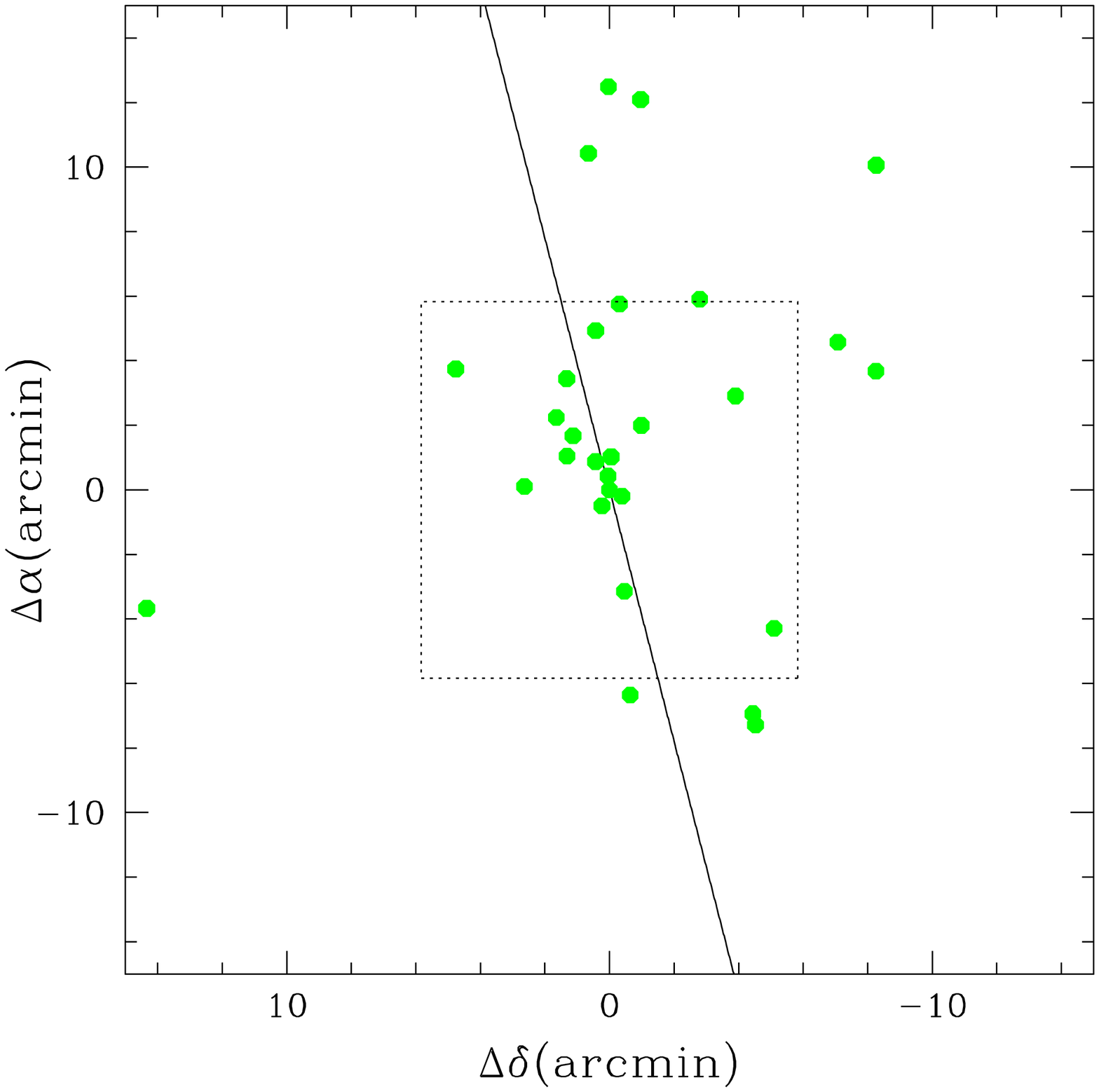}
\end{figure}


\begin{thebibliography}{}
\bibitem[Allen et al. 1995]{all95} Allen, S. W., Fabian, A. C., Edge, A. C., 
	Bohringer, H., \& White, D. A. 1995, \mnras, 275, 741
\bibitem[Andreon, Garilli, \& Maccagni 1995]{and95} Andreon, S., Garilli, B., \& Dario, M. 1995, \aap, 300, 711
\bibitem[Bertin \& Arnouts 1996]{ber96} Bertin, E., \& Arnouts, S. 1996, \aaps, 117, 393
\bibitem[Carlberg et al. 1999]{car99} Carlberg, R. G., Yee, H. K. C., Morris, S. L., Lin, H., 
Ellingson, E., Patton, D., Sawicki, M., \& Shepherd, C. W. 1999, \apj, 516, 522
\bibitem[Carter \& Metcalfe 1980]{car80} Carter, D. \& Metcalfe, N. 1980, \mnras, 191, 325
\bibitem[Cavaliere \& Fusco-Femiano 1976]{cav76} Cavaliere, A. \&
Fusco-Femiano, R. 1976, \aap, 49, 137
\bibitem[Dalcanton 1996]{dal96} Dalcanton, J.J. 1996, \apj, 466, 92
\bibitem[de Vaucouleurs \& de Vaucouleurs 1970]{dev70} de Vaucouleurs,
	G. \& de Vaucouleurs, A. 1970, \apj, 5, L219
\bibitem[Dubinski 1998]{dub98} Dubinski, J. 1998, \apj, 502, 141
\bibitem[Ferreras, Charlot, \& Silk 1999]{fer99} Ferreras, I., Charlot, S., \& Silk, J. 1999, \apj, 521, 81
\bibitem[Fischer 1999]{fis99} Fischer 1999, \aj, 136, 539
\bibitem[Franx, Illingworth, \& Heckman 1989]{fra89} Franx, M., Illingworth, G., \& Heckman, T. 1989, \aj, 98, 538
\bibitem[Garilli et al. 1997]{gar97} Garilli, B., Sangalli, G., Andreon, S., Maccagni, D., 
	Carrasco, L., \& Recillas, E. 1997, \aj, 113, 1973
\bibitem[Girardi et al. 1998]{gir98} Girardi, M., Giuricin, G.,
	Mardirossian, F., Mezzetti, Ml, \& Boschin, W. 1998, \apj, 505, 74
\bibitem[Gonzalez et al. 2000]{gon00} Gonzalez, A. H., Zaritsky, D., Dalcanton, J. J., \& Nelson, A. E.
	2000, in preparation
\bibitem[Graham et al. 1996]{gra96} Graham, A., Lauer, T. R., Colless, M., Postman, M. 
      	1996, \apj, 465, 534
\bibitem[Gregg \& West 1998]{gre98} Gregg, M. D. \& West, M. J. 1998, \nat, 396, 549
\bibitem[Gudehus 1989]{gud89} Gudehus, D. H. 1989, \apj, 340, 661
\bibitem[Jarvis \& Tyson 1981]{jar81} Jarvis, J. F. \& Tyson, J. A. 1981, \aj, 86, 476
\bibitem[Kaufmann \& Charlot 1998]{kau98} Kauffmann, G., \& Charlot, S. 1998, \mnras, 294, 705
\bibitem[Larson 1974]{lar74} Larson, R. B. 1974, \mnras, 166, 585
\bibitem[Mackie 1992]{mac92} Mackie, G. 1992, \apj, 400, 65
\bibitem[Malamuth \& Richstone 1984]{mal84} Malamuth, E. M. \& Richstone, D. O. 1984, \apj, 276, 413
\bibitem[Markevitch et al. 1998]{mark98} Markevitch, M., Forman, W. R., Sarazin, C. L., 
      	\& Vikhlinin, A. 1998, \apj, 503, 77
\bibitem[Marleau \& Simard 1998]{mar98} Marleau, F. \& Simard, L. 1998, \apj, 507, 585
\bibitem[Melnick, Hoessel, \& White 1977]{mel77} Melnick, J., Hoessel, J., \& White, S. D. M. 1977,
        \mnras, 180, 207
\bibitem[Matthews, Morgan, \& Schmidt 1964]{mat64} Matthews, T. A., Morgan, W.
W., \& Schmidt, M. 1964, \aj, 70, 144
\bibitem[Merritt 1984]{mer84} Merritt, D. 1984, \apj, 276, 26
\bibitem[Morgan, Kayser, \& White 1975]{mor75} Morgan, W. W., Kayser, S., White, R. A. 1975, \apj, 199, 545
\bibitem[Mulchaey \& Zabludoff 1998]{mul98} Mulchaey, J. S. \& Zabludoff, A. I. 1998, \apj, 496, 73
\bibitem[Oemler 1973]{oem73} Oemler, A. 1973, \apj, 180, 11
\bibitem[Oemler 1976]{oem76} Oemler, A. 1976, \apj, 209, 693
\bibitem[Porter, Schneider, \& Hoessel 1991]{por91} Porter, A. C., Schneider, D. P., \& Hoessel, J. G. 1991,
	\aj, 101, 1561
\bibitem[Sastry 1968]{sas68} Sastry, G. N. 1968, \pasp, 80, 252
\bibitem[Scheick \& Kuhn 1994]{sch94} Scheick, X. \& Kuhn, J. R. 1994, \apj, 423, 566
\bibitem[Schlegel, Finkbeiner, \& Davis 1998]{schl98} Schlegel, D. J., Finkbeiner, D. P., \& Davis, M. 1998,
	\apj, 500, 525
\bibitem[Schombert 1986]{sch86} Schombert, J. M. 1986, \apjs, 60, 603
\bibitem[Schombert 1987]{sch87} Schombert, J. M. 1986, \apjs, 64, 643
\bibitem[Schombert 1988]{sch88} Schombert, J. M. 1988, \apj, 328, 475
\bibitem[Sersic 1968]{ser68} Sersic, J.-L. 1968, Atlas de Galaxias Australes (Cordoba: Obs. Astronomico)
\bibitem[Simard 1998]{sim98} Simard, L. 1998, Astronomical Data Analysis Software and Systems VII, 
 	A.S.P. Conference Series, Vol. 145,1998, ed. R. Albrecht, R.N. Hook and H.A. Bushouse, 108
\bibitem[Uson, Bough, \& Kuhn 1990, 1991]{uso90} Uson, J. M., Bough, S. P., \& Kuhn, J. R. 1990, Science, 250, 539
\bibitem[Uson, Bough, \& Kuhn 1990, 1991]{uso91} Uson, J. M., Bough, S. P., \& Kuhn, J. R. 1991, \apj, 369, 46
\bibitem[Valdes 1993]{val93} Valdes, F. 1993, FOCAS User's Guide, NOAO document (ftp://iraf.noao.edu/iraf/docs/focas/focasguide.ps.gz)
\bibitem[Vilchez-G\'{o}mez, Pell\'{o}, \& Sanahuja 1994]{vil94}  Vilchez-G\'{o}mez, R., 
Pell\'{o},R., \& Sanahuja, B.  1994, \aap, 283, 37
\bibitem[White et al. 1993]{whi93} White, S. D. M., Navarro, J. F., Evrard, A. E., \& Frenk, C. S. 1993, \nat, 366, 429
\bibitem[Zaritsky, Schectman, \& Bredthauer 1996]{zar96} Zaritsky, D., Shectman, S. A.,
	\& Bredthauer, G. 1996, \pasp, 104, 108 
\bibitem[Zaritsky et al. 1997]{zar97} Zaritsky, D., Nelson, A. E., Dalcanton, J. J., \& Gonzalez, A. H.
	1997, \apj, 480, L91
\bibitem[Zwicky 1933]{zwi33} Zwicky, F. 1933, Helvetica Physica Acta, 6, 110
\end{thebibliography}
\end{document}